\documentclass[preprint,12pt]{elsarticle}

\usepackage{amssymb}

\journal{----}

\begin{document}

\begin{frontmatter}



\title{ Quantum Correlations in Dimers}


\author{$N. Doustimotlagh^{}$\footnote{Corresponding author.\\
\emph{E-mail address}: doustimotlagh@yahoo.com Phone:
+86\,1324\,1699\,374, Fax: +0086(10)62772692}, $^{}$ }

\address{ Tsinghua University, Department of Engineering Physics,
\emph{Beijing 100084}, P.R.China}

\begin{abstract}
It has been proven that the quantum discord is a more general tool to capture non-classical correlation than quantum entanglement, because there is a non-zero quantum discord in several mixed states that could not be measured by quantum entanglement. But because of optimization part in formulating quantum discord, it is very difficult and nearly impossible to find quantum discord for some quantum states. So
people proposed geometric quantum discord, which in a bipartite state could
be describe as the distance of the states from the closest zero-discord state.
To understand better geometric quantum discord, this paper is devoted to compare it with quantum discord and measurement induced non-locality that is in some sense dual to the geometric quantum discord. As studying the quantum correlation experimentally is one
of the important thing in quantum information processing, so as our example for illustrating the difference of geometric quantum discord, quantum discord and measurement induced non-locality, our quantum system is a dimer (two spin-1/2 particles) in multiple quantum nuclear magnetic resonance.

\end{abstract}


\end{frontmatter}


\section{Introduction}
Assume a quantum system that its contents are interacting with each other, one
can describe this system by using the correlations amongst the different constituents of it \cite{Nielson,doosti1}. Generally speaking, correlations have both quantum and classical components. One of the most important type of quantum correlation measures is the quantum entanglement that describes as the basic and essential source in many quantum information processing tasks such as teleportation \cite{hofer2011quantum,yeo}, quantum communication \cite{ursin}, quantum computation \cite{Nielson} and dense coding \cite{yeo}.
It is important to known that quantum mutual information can measure the total
correlation in a bipartite quantum system \cite{Schumacher,Dixon}. Quantum mutual information in turn can be divided into quantum and classical parts and the quantum part is named as quantum discord (QD) that first was described and formulated by the Zurek Group \cite{zurek} and Verdal group \cite{verdal} independently. It has proven that the QD is a more general tool to capture non-classical correlation than quantum entanglement, because there is a non-zero quantum discord in several mixed states that could not be measured by quantum entanglement\cite{Zhang}.
 The formulation of QD is according to the very difficult procedure named numerical maximization and this procedure can not guarantee to calculate exact results and there are analytical expressions for some special cases only \cite{zurek,verdal,hui}. To solve this problem, some physicists proposed another simpler measure called geometric quantum discord (GQD)which in a bipartite state could be describe as the distance of the states from the closest zero-discord state \cite{luo,paula,doosti2}.
To understand better GQD, this paper is devoted to compare it with QD and
measurement induced non-locality (MIN) \cite{luo1} that is in some sense dual to the GQD. As studying the quantum correlation experimentally is one of the important thing in quantum information processing, so we use dimer (two spin-1/2 particles) in multiple quantum nuclear magnetic resonance (MQ NMR)\cite{nmr1,nmr2,nmr3,nmr4,nmr5} to study GQD, QD \cite{nmr5} and MIN numerically.

\section{Theory}
\subsection{Description of State of Dimer in NMR Formalism}
\label{sec:Dimer}

The MQ NMR experiment has four different periods of time that are preparation, evolution, mixing, and detection \cite{nmr1,nmr2,nmr3}. Our system is two spin-1/2 particles with the dynamical decoupling interaction on the preparation period of the MQ NMR experiment which an strong external magnetic field applied on it. We use rotating reference frame \cite{nmr1,nmr2,nmr3} to describe the dynamics of the system. The MQ NMR Hamiltonian $H_{MQ}$ of the system can be writen as \cite{nmr1,nmr2,nmr3,nmr4,nmr5}
\begin{equation}
H_{MQ}=\frac{\eta}2(J^{+}_1J^{+}_2+J^{-}_1J^{-}_2).
\end{equation}
Where $R_{12}$ is the distance between two spins, $\alpha_{12}$ is the angle between the external magnetic field $H_0$ and the vector $R_{12}$. $\eta=(\gamma\hbar/R^3_{12})(1-3cos^2\alpha_{12})$ is the coupling constant between two spins and $\gamma$ is the gyromagnetic ratio. $J^{-}_i$ and $J^{+}_i$ $(i=1,2)$ are the lowering and raising operators of spin $i$. The thermodynamic equilibrium density matrix of the system is
 \begin{equation}
\rho_0=\frac{e^{\beta J_z}}{Tr(e^{\beta J_z})}.
\end{equation}
Where  $\omega=\gamma |H_0|$, $\beta=\frac{\hbar\omega_0}{kT}$ is the inverse temperature and $J_z=J_{1z}+J_{2z}$ describe the z-projection of the total spin angular momentum. Now we can obtain the density matrix, $\rho(\tau)$ \cite{nmr3,nmr4,nmr5}
 
\begin{equation}
\rho(\tau)= \left(
\begin{array}{cccc}
                                            a & 0 & 0 & e \\
                                            0 & b & 0 & 0 \\
                                            0 & 0 & c & 0 \\
                                            -e & 0 & 0 & f \\
                                          \end{array}
                                        \right).
\end{equation}
where
\begin{eqnarray}
&&a=\frac1{z}(\cosh\beta+\cos(\eta t) \sinh\beta),\nonumber\\
 &&e=\frac1{z}(i\sin(\eta t)\sinh\beta ),\nonumber\\
 &&b=c=\frac1{z},\nonumber\\
 &&d=\frac1{z}(\cosh\beta-\cos(\eta t) \sinh\beta).
  \end{eqnarray}

Where $z=2(1+\cosh\beta)$.
\subsection{Quantum Discord in Dimer}
\label{sec:DDisc}

   One can find that $\rho(\tau)$ is a particular case of $X-matrix$ which the QD of such states has been analytically studied \cite{ali}. The QD is the difference of  total mutual and information $I(\rho)$ and the classical correlations $C(\rho)$ \cite{zurek,verdal,nmr5,ali}:
\begin{equation}
QD(\rho)=I(\rho)-C(\rho).
\end{equation}
Where $I(\rho)$ is the total mutual information and it given by \cite{nmr5}
\begin{equation}
I(\rho)=S(\rho^M)+S(\rho^N)+\sum^4_{m=1}{\lambda_m \log_2{\lambda_m}}.
\end{equation}
where $N$ and $M$ denote the second and the first spins respectively. $\rho^M$ and $\rho^N$ are the reduced density matrices and can be obtain as:
\begin{eqnarray}
&&\rho^M=Tr_N\rho=diag(a+b,c+d),\nonumber\\
&&\rho^N=Tr_M\rho=diag(a+c,b+d),\nonumber\\
&&\rho^M=\rho^N\nonumber.\\
&&
\end{eqnarray}
By using $(S(\rho)=Tr(\rho log_2\rho))$ the the appropriate entropies \cite{nmr5} $S(\rho^M)$ and $S(\rho^N)$ can be written as:
\begin{equation}
S(\rho^M)=S(\rho^N)=\frac1{2ln_2}\ln[{\frac{(z/2)^2-\epsilon^2\sinh^2\beta}{z^2}}]-
\frac{\epsilon \sinh\beta}{z\ln_2}\ln[\frac{(z/2)+\epsilon \sinh\beta}{(z/2)-\epsilon \sinh\beta}].
\end{equation}
where $\epsilon=|cos(\eta\tau )|$ introduces the time-dependent parameter and $0\leqslant\epsilon\leqslant1$.  $\lambda_m $s are the eigenvalues of the density matrix and they are independent of $\epsilon$ \cite{ali}:
\begin{equation}
 \lambda_1=\frac{e^\beta}z, \lambda_2=\frac{e^{-\beta}}z,\lambda_3=\lambda_4=\frac1z.
\end{equation}.
We suppose that the projective measurements are applied on the subsystem $N$. The classical correlations $C(\rho)$  \cite{nmr4,nmr5,ali} can be written as:
\begin{equation}
C(\rho)=S(\rho^M)-\min_{\kappa=0,1}\Xi(\kappa,\beta,\epsilon).
\end{equation}
Where
\begin{equation}
\Xi(\kappa,\beta,\epsilon)=p_0f_0-p_1f_1 .
\end{equation}
So one can write\cite{nmr4,nmr5}
\begin{equation}
f_i=-\frac{1-\phi_i}{2}\log_2\frac{1-\phi_i}{2} -\frac{1+\phi_i}{2}\log_2\frac{1+\phi_i}{2}.
\end{equation}
\begin{equation}
p_i=\frac12(1+(-1)^i\kappa(2(a+c)-1)).
\end{equation}
\begin{equation}
\phi_i=\frac1p_i[(1-\kappa^2)|e|^2+\frac14(2(a+b)-1+(-1)^i\kappa(1-2(b+c)))^2]^\frac12.
\end{equation}
Where $\kappa$ is an arbitrary parameter, $0\leqslant\kappa\leqslant1$. One can obtain that the minimum of $\Xi(\kappa,\beta,\epsilon )$ in Eq. (10) is independent of $\epsilon$ and will be correspond to $\kappa = 0$ and after some mathematics calculation one can write:
\begin{equation}
\Xi(\kappa=0,\beta)=f_0=\frac1{2\ln2}[\ln(\frac z4)+\phi_0\ln(\frac {1-\phi_0}{1+\phi_0})]+1.
\end{equation}
where by using Eq. (14) one can obtain $\phi_0=[(\cosh(\beta)-1)/(\cosh(\beta)+1)]^\frac12$.
Now we are able to write a single expression for the QD. By combining Eqs. (11), (12) and (14) and using  $\sum_{m=1}^4\log_2\lambda_m=-2\Xi(0,\beta)$ , one can obtain the following expression for the QD:
\begin{eqnarray}
&&QD(\rho(\tau))=\frac1{2ln_2}ln[{\frac{(z/2)^2-\epsilon^2\sinh^2\beta}{z^2}}]-\frac{\epsilon \sinh\beta}{zln_2}ln[\frac{(z/2)+\epsilon \sinh\beta}{(z/2)-\epsilon \sinh\beta}]\nonumber\\
&&+\frac1{2\ln2}[\ln(\frac z4)+\phi_0\ln(\frac {1-\phi_0}{1+\phi_0})]+1.\nonumber\\
&&
\end{eqnarray}
\subsection{Geometric Quantum  Discord and Measurement Induced Non-locality in Dimer}
\label{sec:DGQD}
By using the density matrix $\rho(\tau)$  we can write GQD for this density matrix as \cite{luo,doosti2}
\begin{equation}
D^g_{A}(\rho(\tau))=\frac14[8(|e|^2+\xi_3-\max(\xi_1,\xi_2,\xi_3)] .
\end{equation}
where
$\xi_1=\xi_2=4(|e|^2)$ and $\xi_3=2[(a-c)^2+(b-d)^2]$.

Locally invariant measurements cause some global effects on quantum systems which one can find quantum correlation in such system by measuring this effects. Luo and Fu \cite{luo1} first proposed a measure to capture these global effect by using a geometric perspective based on the local Von Neumann measurements wherein one of the reduced states is left invariant and called this measure MIN \cite{luo1} that is in some sense dual to the GQD \cite{luo}. Assume that $\rho$ is the density matrix of a bipartite state that act
on the associated state space $H^C \otimes H^C$ and $\dim(H^C \otimes H^C)<\infty$. The MIN of $\rho$ can be defined as \cite{luo1}
\begin{equation}
MIN(\rho)=\max_{\prod^C}\|\rho-\prod^C(\rho)\|^2.
\end{equation}
Where $\|F^2\|=Tr(F^+F)$ is the square of the Hilbert-Schmidt norm and the optimization should be taken over all local Von Neumann measurements $\prod^C=\{\prod^C_k\}$  which should satisfy 
\begin{eqnarray}
&&\sum_k\prod^C_k\rho^C\prod^C_k=\rho^C,\nonumber\\
&& \prod^C(\rho)=\sum_k(\prod^C_k\otimes I_D)\rho(\prod^C_k\otimes I_D).\nonumber\\
&&
\end{eqnarray}
By using the density matrix $\rho(\tau)$ and equations $(5-18)$ we can write MIN in dimer as:
\begin{equation}
MIN((\rho(\tau)))=\frac14[\Gamma+8|e|^2-4\min(\frac\Gamma4,|e|^2)].
\end{equation}
where $\Gamma=(a-b-c+d)^2$.
\section{Numerical Results and Discussion}
\label{sec:Discussion}
To understand better geometric quantum discord, this chapter was devoted to compare it with quantum discord and measurement induced non-locality as a measure of quantum correlation in dimer (two spin-1/2 particles).

\begin{figure}[]
\begin{centering}
\includegraphics[width=14.4cm]{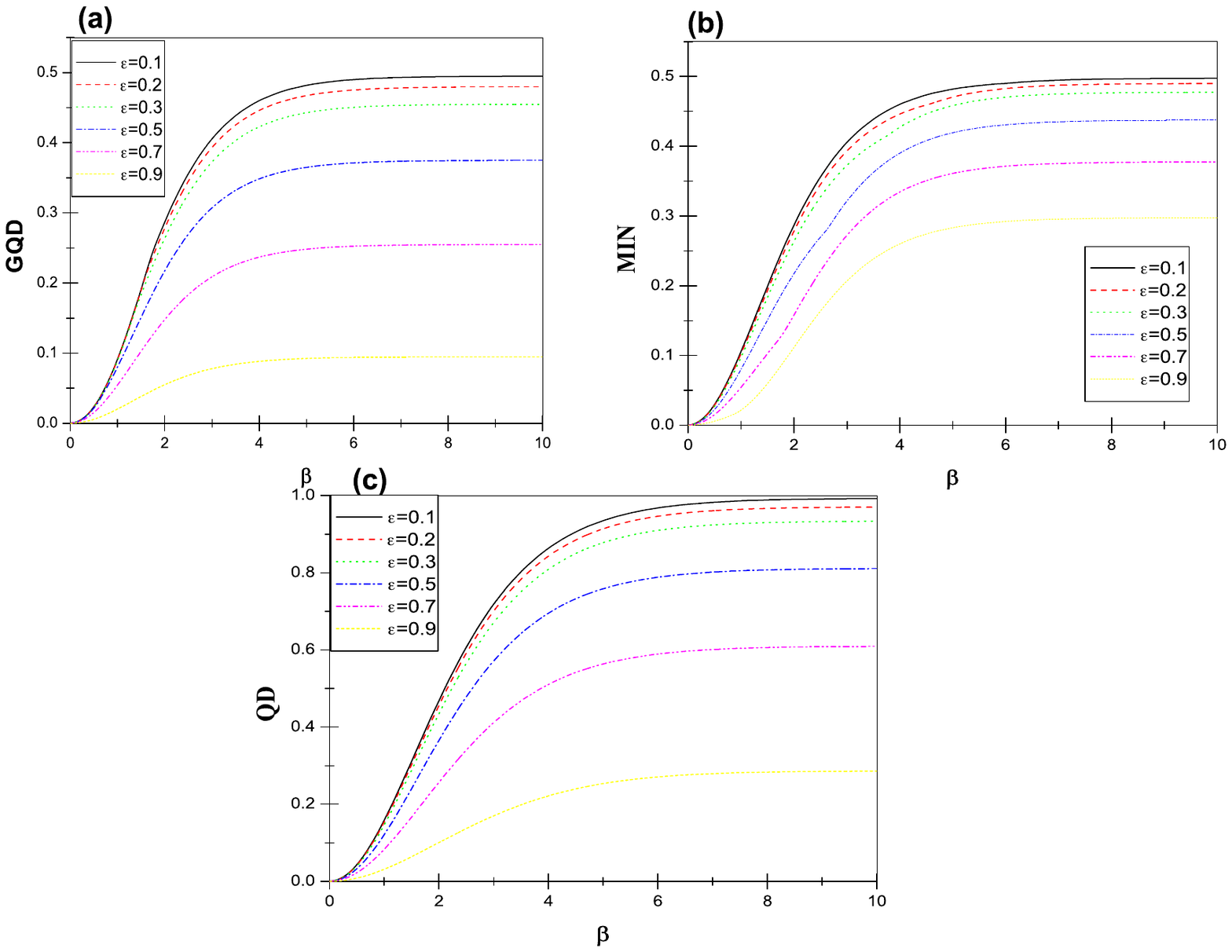}
\caption {GQD, QD and MIN versus $\beta$ for $\epsilon=0.1$ (black and solid), $\epsilon=0.2$ (red and dash), $\epsilon=0.3$ (green and dot), $\epsilon=0.5$ (blue and dash-dot), $\epsilon=0.7$ (magenta and dash-dot-dot) and $\epsilon=0.9$ (yellow and short-dash).}
\label{fig031}
\end{centering}
\end{figure}

\begin{figure}[]
\begin{centering}
\includegraphics[width=14.4cm]{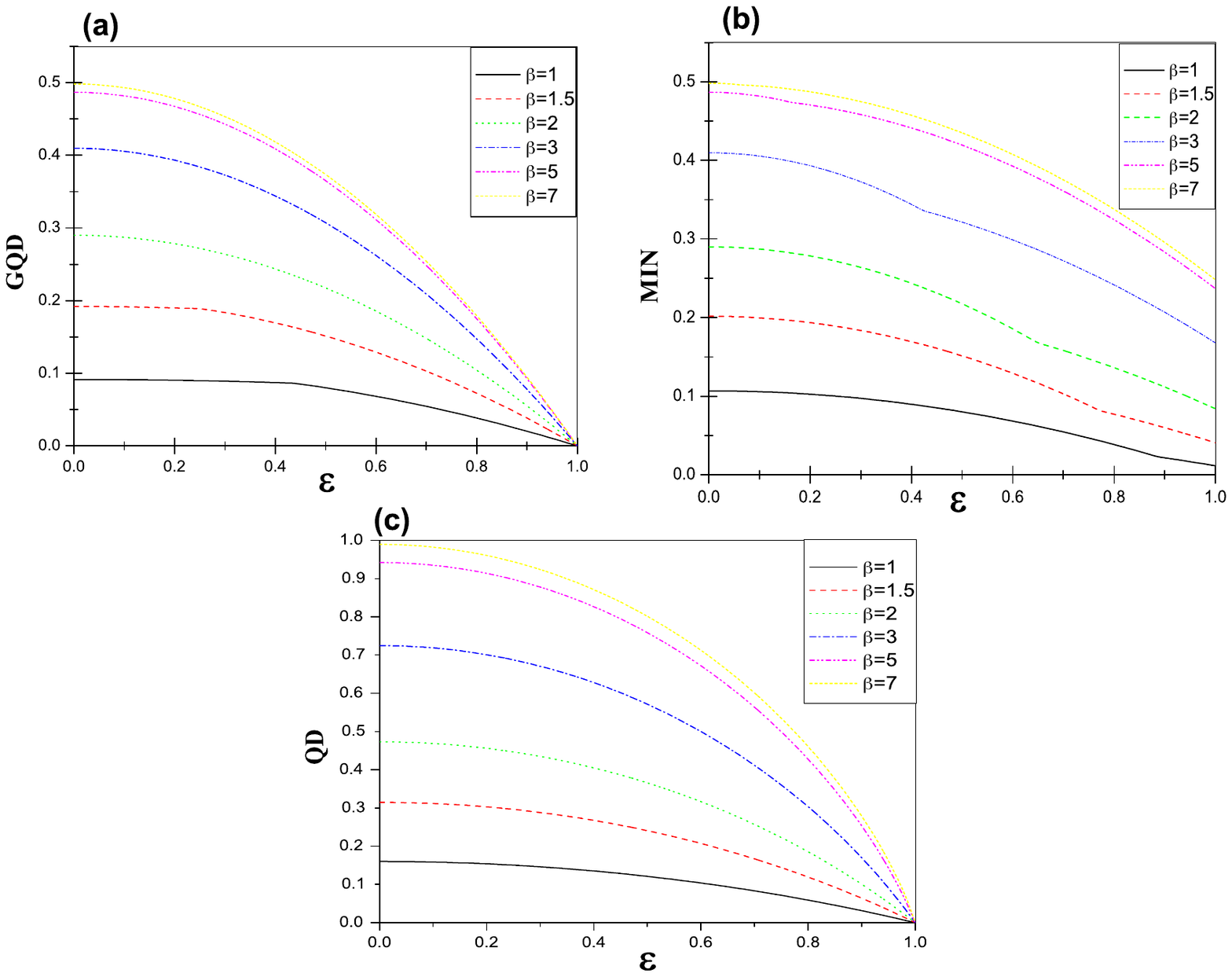}
\caption {GQD, QD and MIN versus $\epsilon$ for $\beta=1$ (black and solid), $\beta=1.5$ (red and dash), $\beta=2$ (green and dot), $\beta=3$ (blue and dash-dot), $\beta=5$ (magenta and dash-dot-dot) and $\beta=7$ (yellow and short-dash).}
\label{fig032}
\end{centering}
\end{figure}

We analyse the GQD, MIN and QD as a function of $\beta$ (reverse of the temperature) for different values of $\epsilon$ (time dependence factor) that one can see our result in Figs. \ref{fig031} wherein (a), (b) and (c) represent GQD, MIN and QD respectively. One can see that for bigger $\beta$ the quantum correlation is bigger, it is natural because $\beta$ is relative to inverse of temperature. As another result one can see that for bigger $\epsilon$
the quantum correlations decrease. It is natural because as the time elapses, interaction of the quantum system with environment will be longer and the quantum correlation must decrease, and to see this result better we draw the GQD, MIN and QD as a function of $\epsilon$ for different values of $\beta$ that one can see our result in Figs. \ref{fig032} wherein (a), (b) and (c) represent GQD, MIN and QD respectively. For understanding the effect of time and temperature simultaneously, we draw the GQD, MIN and QD as a function of $\beta$ and $\epsilon$ in 3-dimensions that one can see our result in Figs. \ref{fig031} wherein (a), (b) and (c) represent GQD, MIN and QD respectively. One cane see that the behaviour of GQD, MIN and QD are nearly same.

\begin{figure}[]
\begin{centering}
\includegraphics[width=17.4cm]{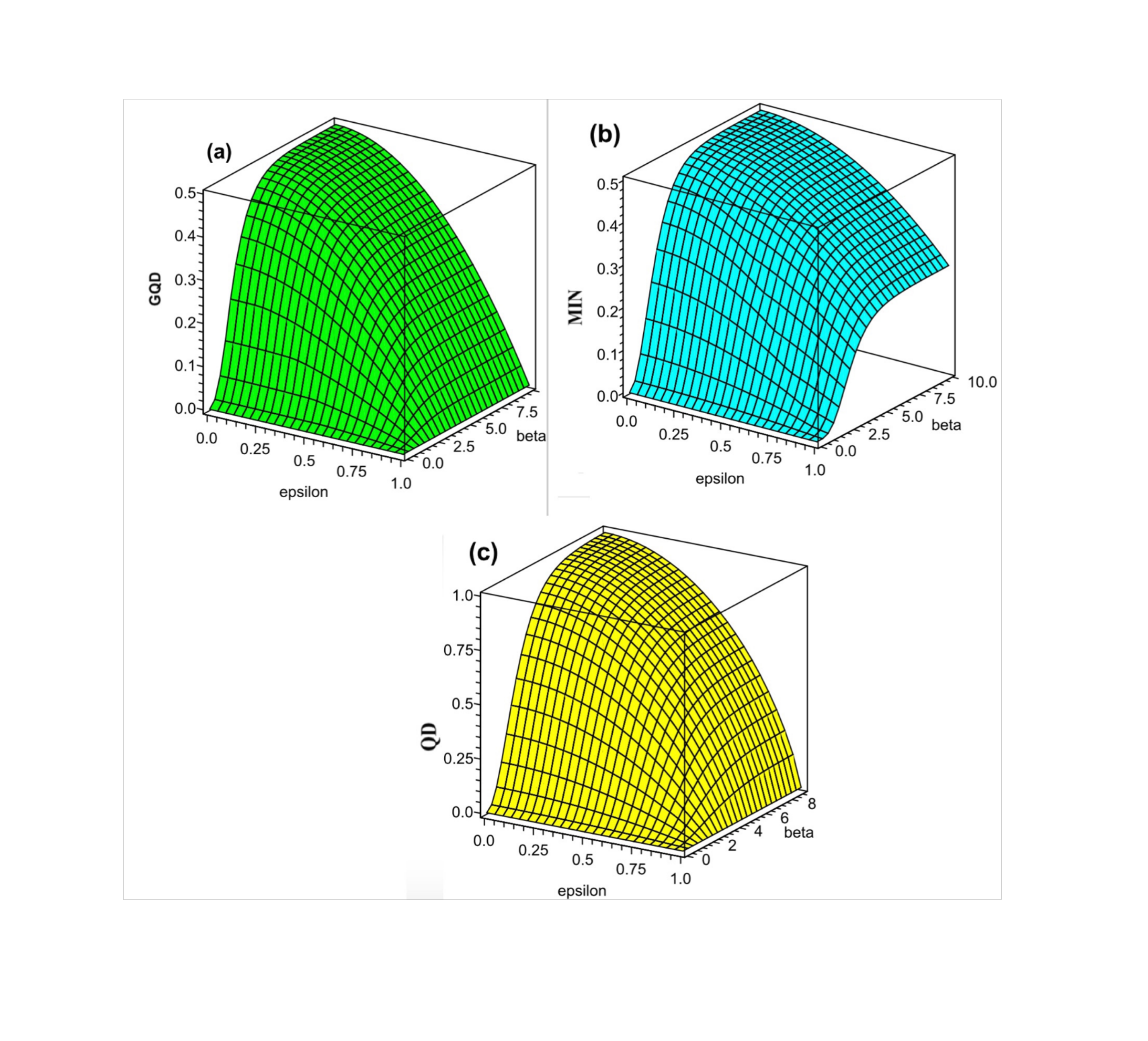}
\caption {(a) GQD, (b) MIN and (c) QD versus $\epsilon$ and $\beta$ in 3-dimensions}
\label{fig033}
\end{centering}
\end{figure}
\section{Summary}
\label{sec:Summary}
 Quantum entanglement as a measure of non-classical correlation plays a central role in the quantum communication and information, however, it could not capture all of the non-classical properties of quantum phenomena. It has proven that the quantum discord is a more general tool to capture non-classical correlation than quantum entanglement, because there is a non-zero quantum discord in several mixed states that could not be measured by quantum entanglement. But because of optimization part in formulating of QD, it is very difficult to formulate quantum discord for some quantum states. So geometric quantum discord, which in a bipartite state could be describe as the distance of the state from the closest zero-discord state was proposed.

To understand better geometric quantum discord, this chapter was devoted to compare it with quantum discord and measurement induced non-locality as a measure of quantum correlation in dimer (two spin-1/2 particles). 
As studding the quantum correlation experimentally is one of the important thing in quantum information processing, in this chapter an scheme for measuring quantum correlation in multi quantum nuclear magnetic resonance was described, wherein Our quantum system is a dimer on the preparation period of this experiment and the measures for quantum correlation is quantum discord, geometric quantum discord and measurement induced non-locality. 
We analyse the quantum discord, geometric quantum discord and measurement induced non-locality as a function of $\beta$ (reverse of the temperature) for different values of $\epsilon$ (time dependence factor) and we found that fore bigger $\beta$ the quantum correlation is bigger, it is natural because $\beta$ is relative to inverse of temperature. As another result, we showed that for bigger $\epsilon$
the quantum correlations decreased. It is natural because as the time elapsed, the interaction of the quantum system with environment will be longer and the quantum correlation must decrease.

\section*{Acknowledgements}
I am so grateful to Professor G. L. Long for his kind help with this work.
\bibliographystyle{model1a-num-names}


\end{document}